# Monitoring Energy Consumption of Smartphones

Fangwei Ding, Feng Xia, Wei Zhang, Xuhai Zhao, Chengchuan Ma
School of Software, Dalian University of Technology, Dalian 116620, China
Email: f.xia@ieee.org

*Abstract*—**With the rapid development of new and innovative applications for mobile devices like smartphones, advances in battery technology have not kept pace with rapidly growing energy demands. Thus energy consumption has become a more and more important issue of mobile devices. To meet the requirements of saving energy, it is critical to monitor and analyze the energy consumption of applications on smartphones. For this purpose, we develop a smart energy monitoring system called SEMO for smartphones using Android operating system. It can profile mobile applications with battery usage information, which is vital for both developers and users.**

## I. INTRODUCTION

With the recent appearance of open operating system and smart phones, new and innovative applications have appeared [1, 2]. For example, from stock tickers to city-wide social games, these devices promise to offer support for a large spectrum of applications [3]. However, many applications such as video-on-demand, mobile gaming, location-aware mobile social applications, and real-time location-based tracking applications [10] are often characterized by heavy network transmission, intensive computation and an always-on display. These features imply a heavy workload on the processors, the wireless network interface and the display in performing these services, which causes a significant energy cost [4].

However, advances in battery technology have not kept pace with rapidly growing energy demands [5, 6, 7]. Most smart phones use rechargeable electrochemical batteries, typically, lithium-ion batteries, as their portable energy source [7]. These fully charged batteries can run on this charge for only a few hours. For example, if the Wi-Fi is used all the time, the smart phone can work for only several hours before it runs out of its energy.

Therefore, the power consumption has emerged as a key issue of the energy management of portables [7, 8, 9]. In this restricted battery environment, developers face new constraints related to the lack of tools for debugging the different aspects that affect mobile energy consumption.

Some existing tools can analyze mobile applications' energy consumption. However, these tools don't address monitoring energy consumption from a developer's standpoint.

To analyze the energy consumption of the applications on mobile phones, we develop a smart energy monitoring system, namely SEMO. It can let developers profile mobile applications with battery information. We develop SEMO software based on Android operating system, which is one of the most popular operating systems for smartphones.

The rest of the paper is organized as follows. Section 2 describes the related work in monitoring energy consumption of mobile devices. We design SEMO, a successful energy consumption monitor for mobile devices and introduce its functions in Section 3. In Section 4, we present prototype implementation of SEMO and some experimental tests, and we conclude the paper in Section 5.

## II. RELATED WORK

Many papers [7, 8, 9, 11] have mentioned that the energy consumption has become an important problem in energy management of mobile phones and have their own ways to save energy. However, we should first know the energy consumption of the applications on mobile phones. Thus, monitoring the energy consumption of smart phones is very important for saving energy to extend the lifetime of battery.

For instance, Crk et al. [12] present a framework for energy monitoring that includes both the hardware and software setup, along with current observations from energy profiling of a mobile phone and a PA monitoring application. However, they focus on understanding the energy consumption of physical activity monitoring applications and do not provide energy consumption monitoring in a developer's perspective. Zhang et al. [13] describe PowerBooter, an automated power model construction technique that uses built-in battery voltage sensors and knowledge of battery discharge behavior to monitor power consumption while explicitly controlling the power management and activity states of individual components. Even though they provide energy consumption monitoring for the developers and users, they do not provide the application-level energy consumption monitoring. However, our SEMO system provides not only energy consumption monitoring for the developers and users, but also the application-level energy consumption monitoring.

## III. SEMO SYSTEM DESIGN

To analyze the energy consumption of the applications on mobile devices, we designed SEMO system. First, it is used to check the battery's status, such as its power remaining and the temperature of its battery. Second, it collects the energy consumption data of the mobile devices, and then it analyzes the energy consumption of the applications on mobile devices according to the data it collects. The collected data include the time, the battery's power remaining at the time and the names of the applications which are running at the time. Third, its data analysis and corresponding algorithms can find the rate of the energy consumption of the applications. It's very useful to the developers and the users of the mobile devices.

As shown in Fig. 1, SEMO consists of the following three main parts: an inspector, a recorder and an analyzer. The inspector is designed to check the information of the battery. The recorder is used to record the information of battery and applications, especially the energy consumption information. Then, the analyzer analyses the data that recorder records to get the rate of the energy consumption of the applications and ranks the applications by these energy consumption rates. In the following sections, we will introduce each part of the SEMO system and explain their functions in detail.

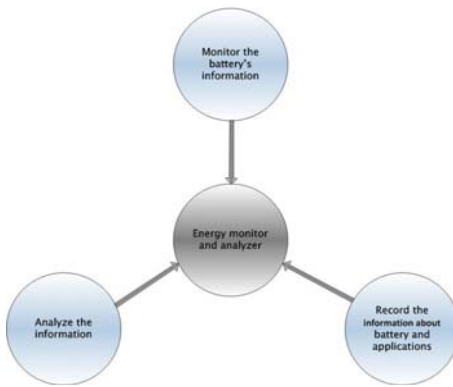

Figure 1. SEMO system structure

### A. The Inspector

We will meet the inspector when using this software. Fig. 2 shows the flow chart of the inspector. As we can see from Fig. 2, the inspector is used to check the information of the battery, and warns users when the battery reaches a critical condition. This is the basic function of most energy monitor. First, it gets the information of the battery, including the percentage of the battery, the health status of the battery, the voltage of the battery, the temperature of the battery and the total battery charge. Then, it warns users when the battery reaches a critical condition. Thus, users can respond appropriately according to the information of the battery. For example, if the percentage of the battery is too lower (such as lower than fifteen), then the inspector will remind users to charge the battery.

### B. The Recorder

The recorder is one of the key parts of the SEMO system, as shown in Fig. 3, which is designed to record the information of battery and the applications on mobile devices, periodically, including the time, the battery's power remaining at the time and the names of the applications which are running at the time. We can change its record interval, and generally, the record interval is set to one minute. With the data of the battery and the applications, we can analyze the energy consumption of the applications later. It can draw the curve of the battery's power remaining changed with the time elapsing in two ways. One is that it draws the history curve of the battery's power remaining according to the information it records. The other is that it draws the real-time curve of the power remaining of the battery.

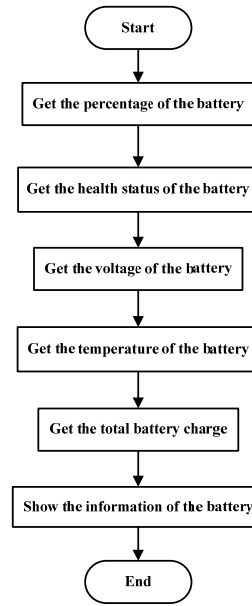

Figure 2. Flow chart of inspector

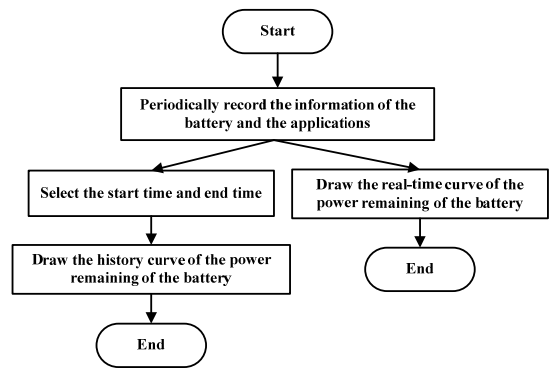

Figure 3. Flow chart of recorder

### C. The Analyzer

The analyzer is also one of the key parts of the SEMO system. Fig. 4 shows the flow chart of the analyzer. As shown in Fig. 4, it is used to analyze the data that recorder records to get the rate of the energy consumption of the applications and rank the applications by their energy consumption rates. In this way, we can find which software cost the most energy on the mobile devices. It will be useful for the user to save energy by closing the most energy consumption application. Or at least, the user can reduce the time of using the application.

Furthermore, we can export the data to the computers with the SQLLITE database if we want a more detailed analysis. We can use the Excel or the Origin software to conduct a more powerful analysis.

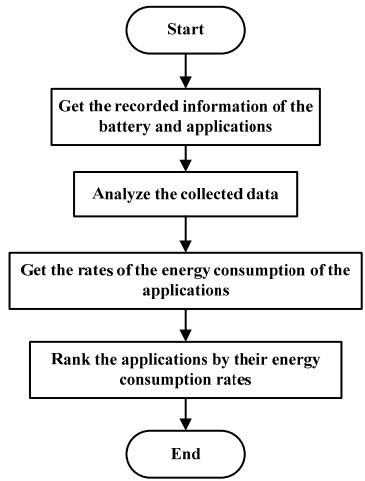

Figure 4. Flow chart of analyzer

The mentioned three major parts in the SEMO system work on the Android system, and its final goal is to get the rate of energy consumption of the applications for developers and save energy for users.

## IV. SEMO PROTOTYPE IMPLEMENTATION

Based on the design described above, we develop a prototype of our SEMO system. Although it is not yet a perfect system, we have implemented the main functions. The prototype is programmed by the Java program language; it is developed on Android SDK (Software Development Kit) [14] by using Eclipse integrated development environment with the ADT (Android Development Tools) Plug-in for Eclipse [15].

We have already tested the SEMO system on both HTC Desire which runs android operating system 2.2 and ZTE-X876 which runs android operating system 2.1. The two devices are shown in Fig. 5. We will take you to have a visit of each part of the SEMO system with some experimental tests in the following sections.

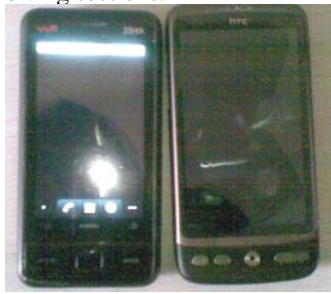

Figure 5. Smartphones with Android OS

### A. Veiw on Inspector

SEMO can work on mobile phones which at least support Android SDK 1.5. After installing it in mobile phones which run android operating system, we can start it and will see the main interface of the SEMO system.

When we click the "Battery Information" icon, we will start the inspector, as shown in Fig. 6. Then, it begin to check the information of the battery, including the percentage of the battery, the health status of the battery, the voltage of the battery, the temperature of the battery and the total battery charge. It warns users when the battery reaches a critical condition.

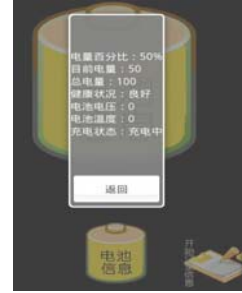

Figure 6. View of inspector

### B. View on Recorder

After successful start the SEMO system, we can click "Start Record Information" icon to start the recorder. Then, it starts to record the information of battery and the applications on mobile devices once per minute, including the time, the battery's power remaining at the time and the names of the applications which are running at the time.

We run the recorder on the ZTE-X876 smart phone which runs text message and web browsing applications except the android operating system. Then, we can draw the curve of the battery's power remaining changed with the time elapsing in two ways, as shown in Fig. 7. One is that it draws the history curve of the battery's remaining power according to the data it records. The other is that it draws the real-time curve of the remaining power of the battery.

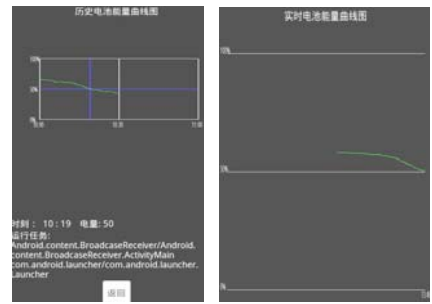

Figure 7. History curve and real-time curve of battery's remaining power

### C. View on Analyzer

After enjoying the recorder, you must be eager to what the analyzer can do. We can start the analyzer by clicking the "Smart Energy Consumption Analysis" icon. Then, it starts to analyze the data that recorder records to get the rate of the energy consumption of the applications and rank the applications by their energy consumption rates.

We use the analyzer to analyze the energy consumption rates of following five tasks (see Table I) and rank them by their energy consumption rates, as shown in Fig. 8. As we can see from the Fig. 8, file download is the most energy-intensive application of the five applications.

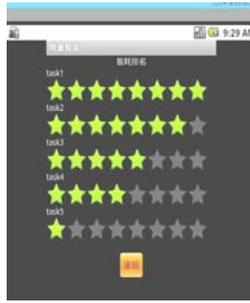

Figure 8. Energy consumption ranking of five tasks

TABLE I. FIVE DIFFERENT TASKS

| Task No. | Applications |
|---|---|
| 1 | File download |
| 2 | Video streaming |
| 3 | Play games |
| 4 | Web browsing |
| 5 | Text message |

## V. CONCLUSIONS

This paper has presented the design of our SEMO system, which can be used to monitor and analyze the energy consumption of applications on smartphones. The software system presented runs on Android operating system, and it is able to accurately record the energy consumption of applications and rank applications according to their energy consumption rates. We have implemented a prototype of our SEMO system and some experimental tests show that it works well on monitoring and analyzing the energy consumption of the applications on smartphones.

ACKNOWLEDGMENTS

This work was partially supported by the Natural Science Foundation of China under Grant No. 60903153, the Fundamental Research Funds for Central Universities (DUT10ZD110), the SRF for ROCS, SEM, and DUT Graduate School (JP201006).